# End-Fire Silicon Optical Phased Array with Half-Wavelength Spacing

**Michael R. Kossey,[1] Charbel Rizk,[1,2] Amy C. Foster[1,*]**

[1]*Department of Electrical and Computer Engineering, Johns Hopkins University, Baltimore, MD 21218, USA*
[2]*Johns Hopkins University Applied Physics Laboratory, Laurel, MD 20723, USA*
*\*Corresponding author: amy.foster@jhu.edu*

*Abstract -* We demonstrate a one-dimensional optical phased array on an integrated silicon platform for operation at 1.55 µm. Light is emitted end-fire from the chip edge where the waveguides are terminated. The innovative design and high confinement afforded by the silicon waveguides enables λ/2 spacing (775-nm pitch) at the output. Steering is achieved by inducing a phase shift between the waveguides via integrated thermo-optic heaters. The device forms a beam with a FWHM angular width of 17°, and we demonstrate beam steering over a 64° range.

A multitude of applications depend on the ability to quickly and accurately steer a laser beam, ranging from fiber endoscopes in biological imaging to LiDAR scanning in large-scale landscape surveying and autonomous automobile navigation. In addition, high-speed beam steering has the potential to impact a host of new technologies from free space optical communications to next-generation projectors. The conventional approaches to realizing optical beam steering are mechanical in nature and require bulky motors and gimbals to physically move the entire laser system, or alternatively tilt-mirrors through galvanometer-based or micromechanical devices. Mechanical systems are fundamentally limited in speed by their inertia. Commercial galvanometer tilt-mirror devices have maximum steering speeds of approximately 10 kHz, whereas the fastest MEMS devices can reach up to 420 kHz [1]. Both types of steering systems can be limited in range as well, with most systems subtending less than 40° total. These limitations have recently prompted the development of non-mechanical beam steering techniques based on integrated optical phased arrays (OPAs) in silicon photonic platforms, which steer by forming a beam from the interference of multiple phase-controlled optical emitters.

The primary physical properties of the phased array that determine its capabilities are the operational wavelength, the spacing of the emission apertures, and the number of apertures [2]. The steering angle will be equal to the angle of the phase front emitted by the array, which, from the ray optics picture, is dependent on the trigonometric relationship between the physical distance covered by phase differences between adjacent apertures and the spacing of those apertures, as described in Equation 1 where $\theta$ is the steering angle, $\varphi$ is the phase shift, and $d$ is the pitch between apertures.

$$\boldsymbol{\phi} = \frac{2\pi d}{\lambda} \, sin(\boldsymbol{\theta}) \qquad (1)$$

The spacing between adjacent apertures also affects the beam pattern formed by the array, with sparser arrays having greater side lobes than a more densely packed array, assuming equal physical area. In addition, the physical area spanned by the array will control the angular size of the main beam, in that a larger array results in a smaller angular divergence. Additionally critical to device performance are the emission properties of individual array elements, which weight the power of the emitted waves.

There are a variety of methods of nonmechanical phase-shifting which can be used in optical phased arrays. Perhaps the most common is the thermo-optic effect. While this is typically on the slower end of nonmechanical steering, usually in the range of 7-200 kHz [3–5], thermo-optic phase shifting can reach speeds of up to 1.5 MHz [6]. Liquid crystal based phase shifters are similar in speed to thermo-optic phase shifters, and have been demonstrated operating at up to 60 kHz [7,8]. While the very best mechanical steering systems are comparable to these previous two nonmechanical methods, electro-optic approaches such as carrier injection can perform phase shifting at much higher, 10's of GHz-rate speeds [9–11].

The majority of OPA demonstrations on integrated silicon platforms use structures such as grating couplers, scattering elements, or nano-antennas that emit light vertically out of the surface of the chip (known as the broadside geometry). These have included some of the largest integrated photonic devices ever developed with arrays consisting of 4096 elements [12]. A variety of devices capable of one-dimensional [13–15] and two-dimensional [16–22] steering have been shown. An alternative geometry for OPAs uses an edge-emitting (end-fire) array rather than vertically-emitting (broadside) geometry [13,15,21]. In the end-fire geometry, light emits into free space in the same plane of the chip via multimode interference couplers or waveguides terminated at the chip edge. This end-fire approach is a fundamentally simpler and lower loss geometry since light remains in the same plane as it exits the chip. While some reflection would occur at the interface between waveguide and free space, reflectance loss for single-mode silicon waveguides is typically below 1 dB. Anti-reflective coatings can reduce this loss further still. In addition, end-fire emitters are extremely compact, which allows small emitter pitch.

Existing silicon integrated OPA designs have two major drawbacks. First, they cannot achieve λ/2 spacing between array elements and are therefore lose power to grating lobes and are limited in their total steering range. Such a small pitch is particularly challenging in broadside approaches where relatively large (with respect to the wavelength) structures such as gratings and nano-antennas cannot physically be spaced λ/2 apart. The second major drawback of current OPAs is that they are not ideal for high-power operation. Grating lobes and vertically emitting structures introduce additional loss to the system. System loss requires a greater amount of input power to achieve a given output power, which increases the chance of damage at the input, where power would be highest. Using materials such as silicon nitride instead of silicon can greatly improve loss performance [23,24], but result in less-confined waveguides, thereby preventing λ/2 spacing.

Achieving λ/2 spacing is ideal for linear phased arrays because it prevents grating lobes: replicas of the central lobe of equivalent magnitude. Grating lobes appear at known angles from the central lobe depending on the aperture spacing of the phased array. The angular range of a phased array is restricted to +/- halfway to the nearest grating lobe, as steering the central lobe beyond this angle will cause aliasing when a grating lobe appears on the opposite side of the steering range. While the range of a uniform phased array with elements spaced greater than λ/2 might be expanded with a diverging lens, it will introduce aberration to the steered beam, particularly near the edges and greatly increase the size of the beam steering system as a whole. Beyond steering range, the presence of additional beams in a beam steering system results in undesirable optical power loss to these unused beams which can be dangerous, and designers must take care to ensure they are blocked. Non-uniform element spacing in phased arrays can reduce the peak amplitude of grating lobes by spreading their energy over a wider area [16,22]. However, this energy does not return to the central lobe, so power is still lost. Finally, smaller element spacing allows a greater number of elements to be packed into a given physical array size, improving resolution and reducing energy in side lobes. Current edge-firing approaches also do not achieve λ/2 spacing due to the use of low refractive index contrast waveguide geometry [21,25], or do not guarantee sufficiently low crosstalk between waveguides to allow small enough spacing [15].

To overcome these drawbacks, we investigate an edge-emitting two-dimensional array of silicon waveguides clad in silicon dioxide. The high index contrast between the silicon core and silicon dioxide cladding of the waveguides results in light being highly confined to the core of the waveguide, thereby allowing dense waveguide spacing at the emission output of the chip at a pitch of λ/2. The low loss design can enable high power density per waveguide, making such an array more suitable for high-power operation. Finally, we can achieve beam steering by integrating active phase shifting in each waveguide i.e. thermo-optic or electro-optic phase shifters. An overview of this scheme is illustrated in Figure 1. Here we demonstrate a one-dimensional (1x5) prototype of the proposed scheme with integrated thermo-optic heaters and achieve a 17° beam width with 64° of steering range. Notably, this work represents the first experimental demonstration of an OPA with λ/2 pitch.

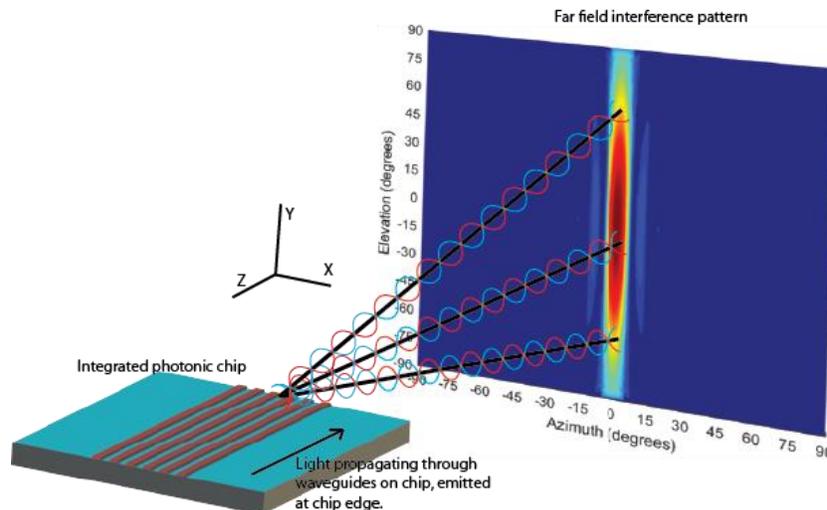

**Fig. 1. Overview of end-firing waveguide optical phased array.**

As shown in our recent work, passive one-dimensional end-fire silicon waveguide arrays are demonstrated as a potential platform for high-speed and high-power beam steering [26]. Though a one-dimensional array's beam forms a stripe rather than a spot and can only steer in one dimension, it can nonetheless prove useful in applications such as one-dimensional imaging. This preliminary generation of devices establishes the ability to form a beam using end-fire waveguides rather than more complicated emission structures. One-dimensional arrays up to 16 waveguides are fabricated in high-confinement silicon waveguides with output spacing of 900 nm. Far-field outputs are measured in the following manner: a 4.5-mm focal length objective lens collected light from the device output and together with a 150-mm focal length convex lens formed a magnified image of the device output. A SWIR camera then captures the far-field pattern formed by this intermediate image. Examples of such far-field patterns are shown in the top of Figure 2 demonstrating the reduction of lobe width as a function of array index. The central lobe FWHM was measured and plotted as a function of array index as shown in the bottom of Figure 2, verifying the expected trend. The 1x16 array of emitters spaced 900 nm apart contains a central beam experimentally measured to be ~7° at FWHM.

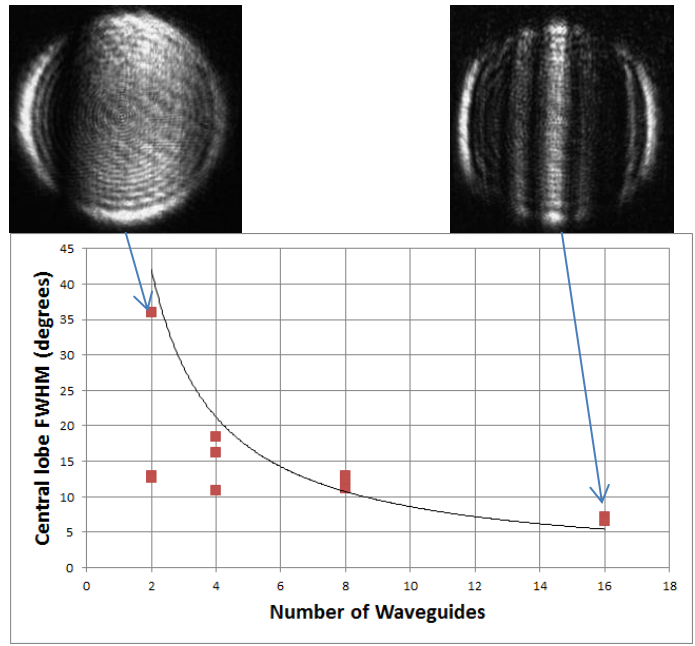

**Fig. 2. FWHM central lobe sizes for our first generation devices compared to simulated results. Squares correspond to measured devices, with the solid line representing simulation. Outliers with 2 and 4 waveguides are due to overpolished outputs, resulting in greater than 900 nm output spacing.**

The second generation optical phased array we present here is a one-dimensional array of five functional elements (the sixth element was not functional). Each element of the array takes the form of a silicon waveguide clad in silicon dioxide, with a width of 500 nm and a height of 220 nm, guiding light with a wavelength of 1550 nm. Similar to our work in [26], the waveguides in this array emit light into free space in the end-fire configuration from terminations at the edge of the chip. We reported preliminary findings in this device in [27]. We calculate the effective refractive index of this waveguide to be 2.445, which results in a reflectance of 0.176 for normal incidence on the interface between waveguide and free space. This corresponds to a reflection loss of -0.84 dB.

As discussed above, $\lambda/2$ element spacing is vital to phased array performance. To determine the smallest value for spacing while keeping crosstalk between waveguides to manageable levels, we utilized the finite difference mode solver from the Photon Design® FIMMPROP™ software to model uniform one-dimensional arrays of waveguides with varying spacing. We found that an array with a waveguide spacing of 775 nm (equal to $\lambda/2$) results in minimal crosstalk (<13 dB) over short propagation distances (10 µm). This, along with crosstalk values for other waveguide spacings, is shown in Fig. 3. It is important to note that this minimum spacing is only required at the output of the device where light is emitted into free space. Elsewhere on the chip the waveguides are routed further apart to mitigate crosstalk. According to our simulations, such an approach can be scaled up to arrays of at least 50 waveguides with dimensions as described above and maintain crosstalk less than 10 dB. Even larger array sizes can be reached by varying waveguide geometry in adjacent elements in order to cause group index mismatch, which has been shown to allow propagation of up to 500 µm at short spacing [28,29].

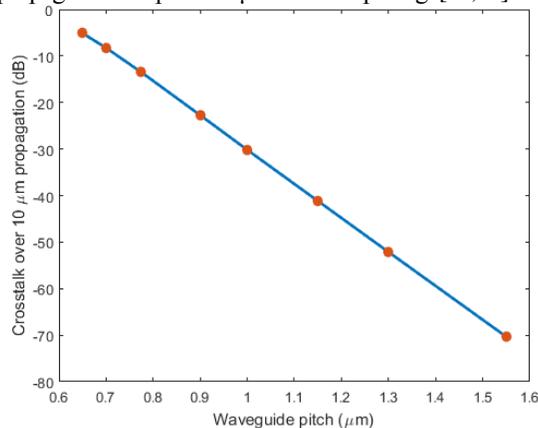

**Fig. 3. Plot of waveguide pitch vs. simulated crosstalk over 10 µm of propagation. For a pitch of 775 nm, half of the operational wavelength of 1550 nm, there will be -13.5 dB of crosstalk.**

We began fabrication with a Soitec® silicon-on-insulator wafer with a 3-μm buried oxide layer. A 100-nm layer of thermal oxide was used as a hard mask to etch the 220 nm thick silicon. We patterned the waveguide array designs via electron beam lithography, etched the hard mask with reactive ion etching and transferred the pattern to the silicon via inductively coupled plasma etching. The cladding is formed by depositing 1 μm of silicon dioxide with plasma enhanced chemical vapor deposition (PECVD). Each waveguide has a total length of 17.8 mm. To achieve steering in our devices, each waveguide has a dedicated thermo-optic phase shifter in the form of a resistive heater that is fabricated as follows: The 1 μm of PECVD device cladding was planarized via chemical mechanical polishing in preparation for fabrication of the thermo-optic phase shifters. A laser writer was used to pattern the heaters followed by a lift-off process. The thermo-optic phase shifters are composed of chromium traces with a thickness of 100 nm, width of 40 μm, and total length of 1.4 mm. The resistivity of this chromium film is $140 \times 10^{-8}$ Ω·m, giving each trace a total resistance of 540 Ω. Thermal simulations show that 10 mA of current applied to one of these traces creates a temperature change of 1.5° Celsius, which for a waveguide length of 5.4 mm in the phase shifter causes a full $2\pi$ phase shift. An image of the finished device is visible in Figure 4, along with CAD and optical microscope images of the phase shifters. All fabrication took place at the NIST Center for Nanoscale Science and Technology NanoFab facility in Gaithersburg, Maryland.

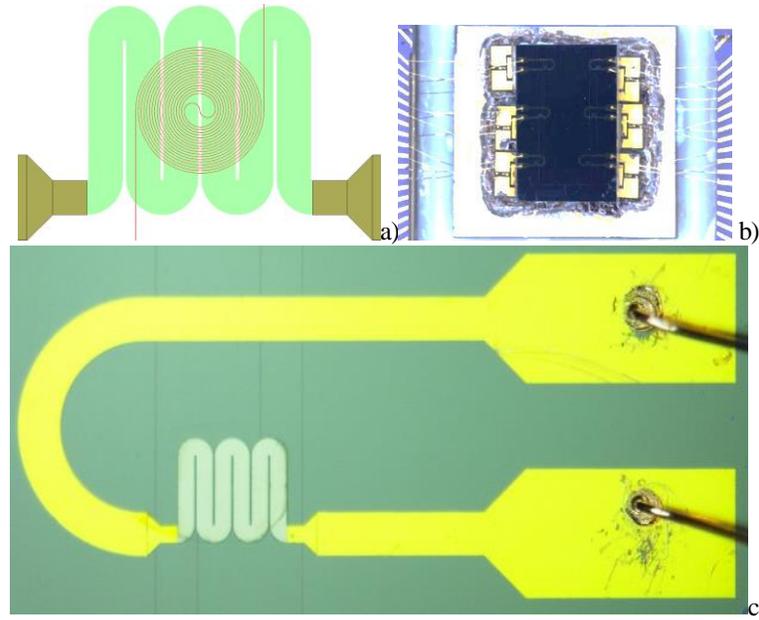

**Fig. 4. a) CAD drawing of thermo-optic phase shifter. b) Image of fabricated device socketed for testing, enhanced to make waveguides visible. c) Optical microscope image of phase shifter after metal deposition and wire bonding.**

Light with a wavelength of 1550 nm from a CW laser was split into five fiber channels and coupled into five of the six waveguides of the phased array device via a Chiral Photonics® Pitch Reduced Optical Fiber Array (PROFA™). The PROFA device has been demonstrated to have <2 dB of insertion loss when butt-coupled to a waveguide edge [30]. The output is imaged using the same imaging setup as described in the proof-of-concept experiment above [26]. The experimental setup is shown in Figure 5. The size, focal length, and numerical aperture of the objective lens in this setup limit the viewable angular range to approximately 60°. Tilting the lens up to 15° off-center while still focusing on the device output enables the extension of the viewing range. By splicing multiple images captured at different angles together, we reach a total measurable range of 90°. We initially image the output of a single waveguide to acquire the element factor, and compare this to the simulated element factor in Figure 6. We see that the simulated element factor has a total width at half maximum of 82°, compared to the approximately 70° shown by the measured element factor. This gives the device an expected total angular steering range of roughly 70°, after which the beam magnitude would fall below half to the amplitude of the beam when centered. We also obtain a simulated vertical element factor, which has a total width at half maximum of 98°.

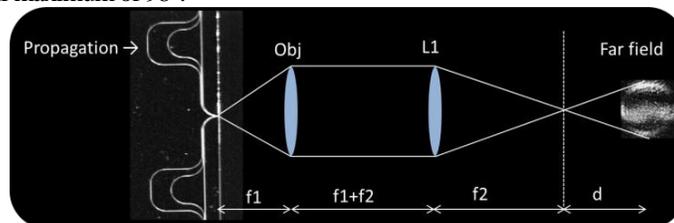

**Fig. 5. Experimental setup. The focal length $f_1$ = 4.5 mm and $f_2$ = 150 mm. The distance from the intermediate image plane that the SWIR camera was positioned $d$ = 135 mm.**

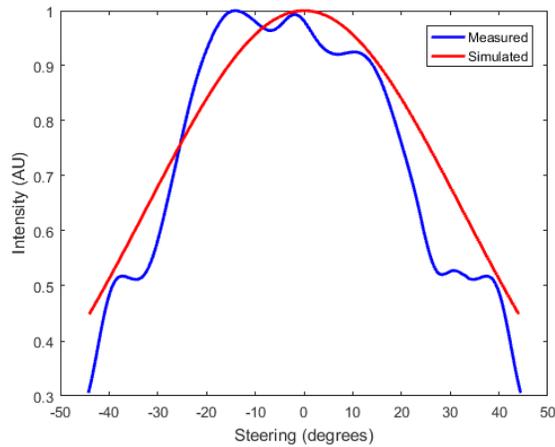
Fig. 6. Simulated and measured element factor for this phased array.

To measure active steering, first we externally equalized the power in each PROFA channel. We set each heater to a starting voltage and then increase and decrease the applied voltage on either side of the array center to create an unwrapped phase ramp at the device output. This resulted in a maximum measured steering of ±32° for a total angular range of 64°, beyond which the steered beam falls below half the peak intensity of the original centered beam, as shown in Fig. 7.

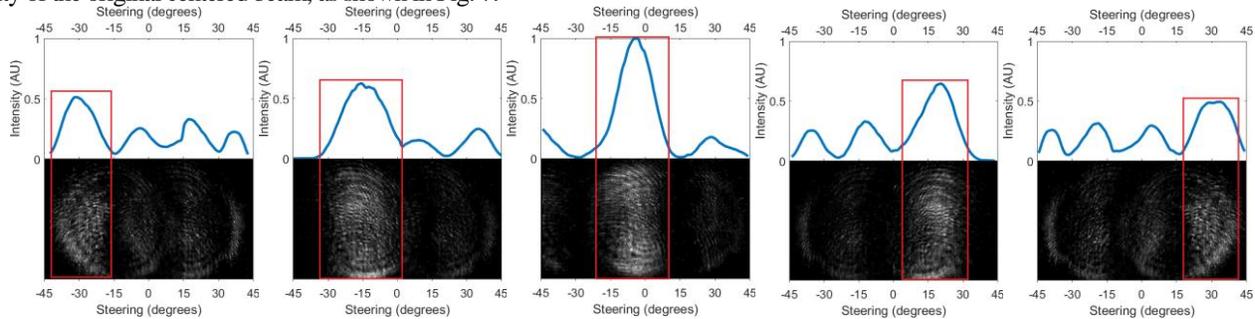
Fig. 7. Central lobe (marked in red) steering from left to right. The top portion of each image shows horizontal cross-section and the bottom shows stitched-together captured images.

Finally, we characterize the $M^2$ beam quality of the central lobe of the interference pattern formed by this device. This is done by focusing the output of the imaging setup, shown in Fig. 5, with a biconvex lens of 100 mm focal length. A slit is used to remove the side lobes from the pattern before focusing. We then capture the focused spot at 5 mm intervals with a SWIR camera and extract the horizontal profile of the beam from the captured image at each step. We fit a Gaussian function to each beam profile, and then calculate the beam FWHM of the obtained curve. We plot the beam widths with their corresponding positions and fit a hyperbolic function to the points. From this function we calculate the beam waist and Rayleigh distance, and subsequently determine the $M^2$ to be 1.20.

Here we demonstrate a one-dimensional OPA device on an integrated silicon photonic platform capable of edge emission of light in the plane of the chip with apertures spaced at λ/2 and steering up to 64°. This shows the large native steering range enabled by using λ/2 element spacing in end-firing waveguide OPAs. Increased resolution can be achieved in future work by increasing the number of waveguide emitters. Furthermore, tailoring the waveguide output geometry can improve element factor, further expanding the steering range. This one-dimensional OPA has potential applications in one-dimensional imaging such as barcode reading, light sheet fluorescent microscopy, and certain LiDAR applications. Three-dimensional integration of these one-dimensional arrays, either by deposition of multiple layers of amorphous materials or by bonding wafers, can enable full two-dimensional phased arrays using this same geometry.